\documentclass{appolb}
\usepackage{amsfonts,epsfig}

\newcommand{\eq}{\begin{equation}}
\newcommand{\eqx}{\end{equation}}
\newcommand{\eqn}{\begin{eqnarray}}
\newcommand{\eqnx}{\end{eqnarray}}
\newcommand{\f}[2]{\frac{#1}{#2}}

\newcommand{\tr}{\mbox{\rm tr}\,}
\newcommand{\Lm}{\Lambda}
\newcommand{\dl}{\delta}
\renewcommand{\th}{\theta}
\newcommand{\thb}{\bar{\theta}}
\newcommand{\Phib}{\tilde{\Phi}}
\newcommand{\al}{\alpha}
\newcommand{\alb}{\bar{\alpha}}
\newcommand{\WW}{{\cal W}}
\newcommand{\ff}{{\cal F}}
\newcommand{\nn}{{\cal N}}
\newcommand{\der}[2]{\f{\partial{#1}}{\partial{#2}}}
\newcommand{\Qt}{\tilde{Q}}

\newcommand{\OM}{\Omega}
\newcommand{\upnc}{{\cal U}^{pure}}
\newcommand{\upm}{{\cal U}^{matter}}
\newcommand{\uf}{u^{fact.}}
\newcommand{\cor}[1]{\left\langle {#1} \right\rangle}
\newcommand{\qqqq}{\quad\quad\quad\quad}

\newcommand{\bin}[2]{\left(\!\!%
\begin{tabular}{c}
{$#1$}\\
{$#2$}
\end{tabular}
\!\!\right)
}

\title{The Dijkgraaf-Vafa correspondence for theories with fundamental
  matter fields\thanks{Talk presented at the Workshop on {\em 
Random Geometry}, Krak\'{o}w, May 15th-May17th, 2003}}

\author{Romuald A. Janik\address{
 M. Smoluchowski Institute  of Physics,\\ 
Jagellonian University,\\ 
Reymonta 4,\\
30-059 Krak\'{o}w, Poland}
}

\begin{document}

\maketitle

\begin{abstract}
In this talk I describe some applications of random matrix models to
the study of $\nn=1$ supersymmetric Yang-Mills theories with matter
fields in the fundamental representation. I review the derivation of
the Veneziano-Yankielowicz-Taylor/Affleck-Dine-Seiberg superpotentials
from constrained random matrix models (hep-th/0211082), a field theoretical
justification of the {\em logarithmic} matter contribution to the
Veneziano-Yankie\-lowicz-Taylor superpotential (hep-th/0306242) and 
the random matrix based solution of the complete factorization problem
of Seiberg-Witten curves for $\nn=2$ theories with fundamental matter
(hep-th/0212212).  
\end{abstract}

\pagestyle{plain}

\section{Introduction}

One of the most important and challenging problems in physics is the
understanding of the non-perturbative properties of gauge theories.
In the last 10 years a series of significant breakthroughs were made
in the study of {\em supersymmetric} gauge theories, where the
additional symmetry properties allowed to obtain exact non-perturbative
information at the same time unraveling various unexpected
mathematical structures.

The first wave of research was based on exploiting the holomorphic
properties of various $\nn=1$ non-perturbative quantities (see
e.g. \cite{SEIBHOL}). The second breakthrough, this time giving very
complete information on $\nn=2$ theories, showed that the low energy
dynamics is encoded in properties of certain elliptic (or
hyperelliptic) curves --- the Seiberg-Witten curves \cite{SW1,SW2}.

The most recent breakthrough arose in the wake of the gauge
theory/string correspondence \cite{adscft}. Using D-brane
constructions of gauge theories and performing `geometric transitions'
to a dual geometric setup, superpotentials in supersymmetric gauge
theories where related to the geometry of the dual Calabi-Yau manifold
\cite{V1,V2}. In a subsequent crucial development Dijkgraaf and Vafa 
interpreted the geometrical formulas in terms of a saddle point
solution of an associated random matrix model
\cite{DV1,DVP}. Subsequently the resulting random matrix prescription for
calculating the superpotential was proven purely in the
field-theoretical context by using diagrammatic analysis
\cite{DVZANON} and anomaly considerations \cite{CDSW}.

In this talk, after recalling some basic facts about $\nn=1$ SYM
theories, I would like to review three developments \cite{DJ1,AJ,DJ2} 
in the extension of the original Dijkgraaf-Vafa proposal to
supersymmetric gauge theories with matter fields in the fundamental
representation. 

The original proposal involved only superpotentials
expressed in terms of the glueball (gaugino bilinear) superfield
$S$. In \cite{DJ1} we showed 
how one could incorporate into the framework mesonic superfields, and
how to obtain the classical
Veneziano-Yankielowicz-Taylor/Affleck-Dine-Seiberg superpotentials
directly from the matrix model. 

The structure of the superpotentials involve typically logarithmic and
power-series terms. The latter are quite well understood
\cite{DVZANON,CDSW} while the former remain somewhat mysterious.
In \cite{AJ} we gave a field-theoretic
diagrammatic derivation of the {\em matter induced} logarithmic part
of the VYT potential.

Finally I would like to review the use of random matrix models to
obtain an explicit complete factorization \cite{DJ2} of the
Seiberg-Witten curve of an $\nn=2$ $U(N_c)$ SYM theory with $N_f<N_c$
fundamental flavours. This is equivalent to finding the submanifold of
$\nn=2$ moduli space of vacua where all monopoles in the theory become
massless. 

\section{Physics of $\nn=1$ SYM theories with fundamental matter}

In order to study the dynamics of the theory at low energies one is
interested in obtaining the low energy effective action for the
relevant degrees of freedom. In the case of $\nn=1$ gauge theories 
most of such
degrees of freedom may be described by chiral (and anti-chiral)
superfields. Then the constraints of supersymmetry restrict the
effective action to be of the form:
\eq
S_{eff} = \int d^2 \th d^2 \thb\, K(\Phi,\Phib) + \int d^2\th
\,W_{eff}(\Phi) +\int d^2\thb\, \tilde{W}_{eff}(\Phib) 
\eqx
where $\Phi$ stands here for a generic chiral superfield.
In general little is known about the non-chiral part $K(\Phi,\Phib)$,
but a lot of information can be obtained on the effective
superpotential $W_{eff}(\Phi)$. Its knowledge allows us to determine
vacuum expectation values (VEV) of $\Phi$ from the equation
\eq
\der{W_{eff}(\Phi)}{\Phi} =0
\eqx
Let us briefly recall what is known about effective superpotentials
first for  pure $\nn=1$ SYM with gauge group $SU(N_c)$, and then for
the theory with $N_f$ fundamental flavours.

\subsubsection*{Pure $\nn=1$ SYM}

A natural chiral superfield which can be built from the gauge field is 
\eq
\label{e.sdef}
S=-\f{1}{32 \pi^2} \tr \WW_\al \WW^\al
\eqx 
whose lowest component is the gaugino bilinear. The form of the
effective superpotential $W_{eff}(S)$ has been determined by Veneziano
and Yankielowicz based on anomaly considerations \cite{VY}:
\eq
W_{VY}(S)=- S \log \f{S^{N_c}}{\Lm^{3N_c}} +2 \pi i \tau_0 S
\eqx
The second piece is just the tree level YM coupling
($\tau_0=\theta/2\pi +4\pi i/g^2_{YM}$) and can be
absorbed into the first one through a redefinition of $\Lm$. The power  
$S^{N_c}$ is determined by the anomaly, while the exponent of $\Lm$ is
just the coefficient of the 1-loop $\beta$ function, in order for the
superpotential to be RG invariant. $W_{VY}(S)$ leads to a nonzero VEV
for S --- gaugino condensation.

\subsubsection*{$\nn=1$ SYM with $N_f<N_c$ flavours}

When we add to the theory matter fields in the fundamental
representation there appear additional mesonic chiral superfields:
\eq
X_{ij}=\Qt_i Q_j
\eqx
Veneziano, Yankielowicz and Taylor determined the appropriate
superpotential again using anomaly considerations \cite{VYT}
\eq
W_{VYT}(S,X)= (N_f-N_c) S \log \f{S}{\Lm^3}-S \log\f{\det X}{\Lm^2}
=-S \log \left( \f{S^{N_c-N_f} \det X}{\Lm^{3N_c-N_f}} \right)
\eqx
Note the modification of the anomaly coefficient ($N_c-N_f$) and the
1-loop $\beta$ function $3N_c-N_f$ due to the matter fields. Later we
will show how these modifications arise from the matrix model framework.
After one integrates out $S$ one is left with the Affleck-Dine-Seiberg
superpotential \cite{ADS} for the mesonic superfield only:
\eq
\label{e.ads}
W_{ADS}(X)=(N_c-N_f) \left( \f{\Lm^{3N_c-N_f}}{\det X}
\right)^{\f{1}{N_c-N_f}} 
\eqx

\subsubsection*{Intrilligator-Leigh-Seiberg (ILS) linearity principle}

Suppose that the gauge theory is deformed by some tree level potential
$V_{tree}(X)$ (e.g. mass terms for the flavours and some
self-interactions). Intrilligator, Leigh and Seiberg \cite{ILS} conjectured
that the full dynamics is described just by the sum
\eq
W_{VYT}(S,X)+V_{tree}(X)
\eqx  
i.e. the dynamical potential is not modified by the deformation.

This concludes the very brief review of classical results on $\nn=1$
SYM theories. In the next section we will present the main new
ingredient --- the link with random matrix models proposed by Dijkgraaf
and Vafa, and proceed to describe its applications.  

\section{The Dijkgraaf-Vafa correspondence}

Dijkgraaf and Vafa proposed an effective way of calculating
superpotentials for $\nn=1$ theories with adjoint fields and arbitrary
tree level superpotentials \cite{DVP} (later generalized also to include
fields in the fundamental representation \cite{FERR1}):
\eq
\label{e.prop}
W_{eff}(S)=W_{VY}(S) + N_c \der{\ff_{\chi=2}(S)}{S}+\ff_{\chi=1}(S)
\eqx
where the $\ff_i$'s are defined through a matrix integral
\eq
\label{e.presc}
e^{-\sum_{\chi} \f{1}{g_s^\chi} \ff_\chi (S)} =
\int D\Phi DQ_i D\Qt_i \, \exp\left\{ -\f{1}{g_s}
W_{tree}(\Phi,Q_i,\Qt_i) \right\}
\eqx
Here $\Phi$ is an $N\times N$ matrix, while the $Q_i$'s are $N$
component (complex) vectors. In this expression one takes the limit
$N\to \infty$, $g_s \to 0$ with $g_s N=S=const$ in order to isolate
graphs with the topology of a sphere ($\ff_{\chi=2}$) and of a disk
($\ff_{\chi=1}$).

The above expression gives a prescription for the
superpotential only in terms of the glueball superfield $S$. However
it is interesting to ask if one could obtain directly from matrix
models the superpotential which involves also mesonic
superfields. A proposal for doing that was given in \cite{DJ1} and will be
described in the next section. 

\section{Mesonic superpotentials from Wishart random matrices}

In order to express the effective superpotential in terms of
mesonic superfields $X_{ij}=Q^\dagger_i \Qt_j$, it was proposed in \cite{DJ1}
to perform only a partial integration over the $Q's$ in 
(\ref{e.presc}) and impose the constraint $X_{ij}=Q^\dagger_i \Qt_j$
directly in the matrix integral i.e.
\eq
\label{e.modpresc}
e^{-\sum_\chi \left(\f{N}{S}\right)^{\chi} \ff_\chi (S,X)} =
\int  DQ_i D\Qt_i \, \dl(X_{ij}- Q_i^\dagger \Qt_j) \exp\left\{ -\f{N}{S}
V_{tree}(X) \right\}
\eqx
Then the effective superpotential involving both $S$ and $X$ is
obtained from 
\eq
W_{eff}(S,X)=W_{VY}(S)+ N_c \der{\ff_{\chi=2}(S,X)}{S}+\ff_{\chi=1}(S,X)
\eqx
For theories with matter fields only in the fundamental representation
(as written in (\ref{e.modpresc})) the above simplifies since
$\ff_{\chi=2}=0$. 

From (\ref{e.modpresc}) we see that $V_{tree}(X)$ contributes directly
to $\ff_{\chi=1}(S,X)$. This is in complete agreement with the ILS
principle where the `dynamical' Veneziano-Yankielowicz-Taylor
superpotential is not influenced by the tree level deformation.

The `dynamical' contribution to $\ff_{\chi=1}(S,X)$
will come from the constrained integral over $N_f$ vectors of length $N$
\eq
\label{e.meas}
\int  DQ_i D\Qt_i \, \dl(X_{ij}- Q^\dagger_i \Qt_j)
\eqx
Up to an inessential term $\exp(-\tr X)$ this is just the probablity
distribution of (complex) Wishart random matrices \cite{WISHART}.
The result is known\footnote{See e.g. \cite{US} for
a general proof and eq (15) in \cite{FYODOROV} for the numerical
coefficient.} and for the case $N>N_f$ reads
\eq
\int DQ D\Qt\, \dl(\Qt Q-X)=
\f{(2\pi)^\f{N(N+1)}{2}}{\prod_{j=N-N_f+1}^N (j-1)!} \left( \det X
\right)^{N-N_f} 
\eqx 
Extracting now the large $N$ asymptotics according to
(\ref{e.modpresc}) one sees that the normalization factor gives rise
to a $N_f S \log S$ term, which reflects the contribution of the matter
fields to the anomaly. The $\det X$ term leads to the correct
dependence on the mesonic superfield.
The full superpotential is then
\eq
\label{e.weffsx}
W_{VY}(S)+ N_f S \log \f{S}{\Lm^3} -S \log\left(\f{\det
X}{\Lm^{2N_f}}\right)   +V_{tree}(X)=W_{VYT}+V_{tree}(X)
\eqx 
as expected from anomaly considerations \cite{VYT}. It is quite surprising
that both terms arise from the classical random matrix Wishart
distribution. 

\subsubsection*{The case with $N_f=N_c$}

It is interesting to consider the case $N_f=N_c$ with $V_{tree}(X)=0$. From
the general form of (\ref{e.weffsx}) we see that the logarithmic term
vanishes and $S$ appears only {\em linearly} as
\eq
W_{eff}=S\log \left(\det X/\Lm^{2N_c} \right)
\eqx
thus it generates in a natural way a constraint surface (moduli space)
satisfying the
equation $\det X =\Lm^{2N_c}$. On this surface the effective potential
vanishes. This is the correct behaviour for $N_f=N_c$
and agrees with Seiberg's quantum constraint 
\eq
\det X -B \bar{B} =\Lm^{2N_c}
\eqx
when the baryonic fields are integrated out \cite{SEIBERG}.
However a direct inclusion of the baryonic superfields into the random
matrix or a more general combinatorial framework remains an open
problem \cite{ROIBANBAR,FERR2}.

\section{Perturbative justification of mesonic superpotentials}

For adjoint matter fields the non-logarithmic part of the Dijkgraaf-Vafa
prescription has been verified by a perturbative diagrammatic
calculation \cite{DVZANON} and arguments based on generalized Konishi
anomalies \cite{CDSW}. However the origin of the logarithmic
Veneziano-Yankielowicz term remains from this point of view quite
obscure --- as it should arise from Feynman graphs which involve
`dynamical' vector superfield loops.

Since a logarithmic term also appeared from a random matrix
calculation in the constrained matrix integral it is interesting to
check if the matter induced part of the Veneziano-Yankielowicz
superpotential may be also obtained diagramatically directly in the
gauge theory. This calculation was done in \cite{AJ}.

The matter contribution to the effective potential in $S$ (with all matter
superfields integrated out) is given by \cite{DVZANON}
\eq
\label{e.pathint}
\int DQ D\Qt \; e^{ \int d^4 x d^2\th \; 
\left(-\f{1}{2} \Qt( \Box - i \WW^\al
\partial_\al) Q +W_{tree}(\Qt,Q)\right) }
\eqx
where $\WW^\al$ is the external field related to $S$ through
(\ref{e.sdef}) and $\partial_\al \equiv
\f{\partial}{\partial\th^\al}$. In \cite{AJ}, in order to obtain the
effective superpotential involving also mesonic superfields we
introduce the superspace constraint
\eq
X=\Qt Q
\eqx
by inserting into (\ref{e.pathint}) a Lagrange multiplier chiral
superfield $\al$. Using the fact, exploited in \cite{DVZANON}, that
the antichiral 
sector does not influence chiral superpotentials we introduce also an
antichiral  partner $\alb$ with a tree
level potential $\alb^2$ i.e.
\eq
\int d^4x d^4\th \;\alb \al +\int d^4x d^2\th \;  \alb^2.
\eqx
The path integral over $\alb$ can be carried out exactly and yields
(c.f. \cite{DVZANON})
\eq
-\f{1}{2}\int d^4x d^2\th \;{\al \Box \al}.
\eqx
Therefore the final path integral which one has to evaluate is 
\eq
\label{e.pathintfin}
\int D\al D\Qt DQ \; e^{\int d^4 x d^2\th 
\left(-\f{1}{2} \Qt( \Box - i \WW^\al
\partial_\al) Q - \f{1}{2} \al \Box \al - \al X +\al \Qt Q\right)},
\eqx
This is a nontrivial interacting field theory, but since we want to
extract only the $\tr \WW^2$ terms we can allow only at most two $\WW$
insertions in a $\Qt Q$ loop. The structure of the integration over
fermionic momenta then significantly reduces the number of
contributing graphs. In particular we are left only with graphs coming
from (\ref{e.pathintfin}) which have the structure of $\Qt Q$ loops
connected by at most one $\al$ propagator and $\al$ propagators
connected to the external field $X$ as shown in fig. 1. Since the
$\al$ propagators are then evaluated necessarily at zero momentum one
has to include an IR cut-off $\Lm_{IR}$.

\begin{figure}
\centerline{\epsfysize=3cm  \epsfbox{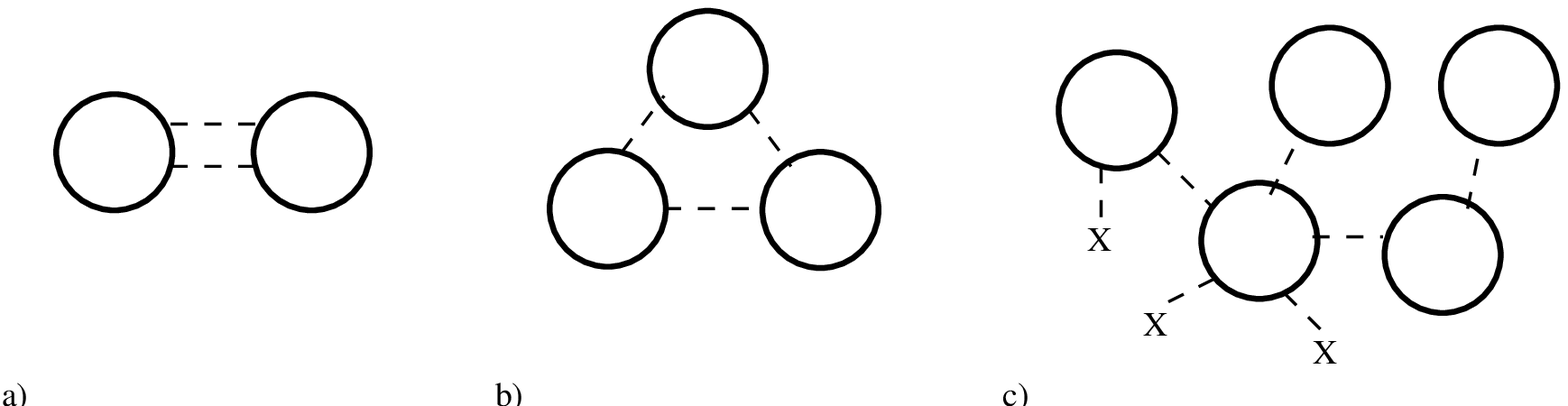}}
\caption{Only tree level graphs survive (like the one shown in fig.\
  \ref{fig1}c).} 
\label{fig1}
\end{figure}

\begin{figure}
\centerline{\epsfysize=1cm  \epsfbox{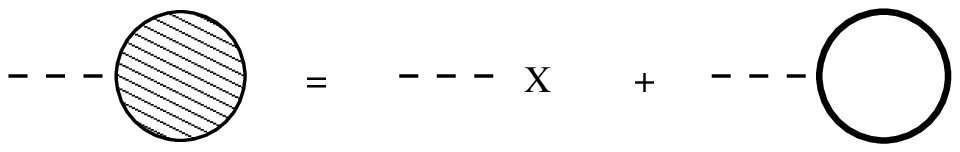}}
\caption{The Schwinger-Dyson equation for $F$.}
\label{fig2}
\end{figure}

The form of the superpotential then follows from (for details see \cite{AJ})
\eq
\label{e.sf}
S \log \det \f{F}{\Lm}
\eqx
where $F$ is the $\al$ 1-point function determined by a
Schwinger-Dyson equation (fig. 2):
\eq
\label{e.sd}
F=-\f{1}{\Lm_{IR}} X +\f{1}{\Lm_{IR}} \,\f{S}{F}
\eqx
The only IR-finite solution is $F=S X^{-1}$ which when inserted into
(\ref{e.sf}) yields the expected result (compare with
(\ref{e.weffsx})) 
\eq
N_f S \log \f{S}{\Lm^3} -S\log \det \f{X}{\Lm^2}
\eqx
including the matter generated logarithmic term.

\section{Random matrices and Seiberg-Witten curves}

In the preceeding sections we described the rederivation using either
random matrix or diagrammatic methods of the
Veneziano-Yankielowicz-Taylor/Affleck-Dine-Seiberg
superpotentials. It is interesting to ask whether one can use the
close relation between random matrix models and SYM theories with
matter fields
to obtain some new nontrivial information about the latter. An example
of this kind of result is the effective factorization of
Seiberg-Witten curves from random matrix considerations \cite{DJ2}.
Before we describe it in the following section let us first recall
some basic facts about $\nn=2$ SYM theories.

At low energies $\nn=2$  $U(N_c)$ SYM theories develop an
$N_c$-dimensional moduli space of vacua parametrized by vacuum
expectation values
\eq
u_p =\cor{\f{1}{p} \tr \Phi^p}
\eqx
with $1 \leq p \leq N_c$.
The IR dynamics of the theory around each such vacuum is described by the
Seiberg-Witten curve
\eq
\label{e.swgen}
y^2=P_{N_c}(x,u_k)^2 -4 \Lm^{2N_c-N_f} \prod_{i=1}^{N_f} (x+m_i)
\eqx
where the polynomial $P_{N_c}(x,u_k)=\cor{\det(xI-\Phi)}\equiv
\sum_{\al=0}^{N_c} s_\al x^{N_c-\al}$ depends in an
explicit way on the $u_k$'s:
\eqn
&&\al s_\al +\sum_{k=0}^\al k s_{\al-k} u_k =0 \\
&&s_0=1, \qqqq u_0=0
\eqnx
Of particular interest (cf. \cite{DS}) are the vacua where all the
monopoles of the theory are massless. Then the Seiberg-Witten curve
{\em completely factorizes}, i.e. the right hand side of
(\ref{e.swgen}) has only {\em two} single zeroes, all the remaining
zeroes are double. The Seiberg-Witten curve then has to have the form
\eq
y^2=(x-a)(x-b) H_{N_c-1}(x)^2
\eqx
The problem is to find the moduli $\{u_k\}$'s for which this happens.  
One expects a 1-parameter family of solutions.
For pure $\nn=2$ theory ($N_f=0$) the explicit form of this
submanifold of vacua has been found using special properties of
Chebyshev polynomials \cite{DS}. Unfortunately these methods cannot be
extended to the case of
$N_f>0$. In \cite{DJ2}, using random matrix model techniques, we
obtained formulas for $u_k$ when the general form of the Seiberg-Witten curve
(\ref{e.swgen}) completely factorizes for $N_f<N_c$ and arbitrary
masses $m_i$.  

We note that the general problem of factorizing (\ref{e.swgen}) is
highly nonlinear and involves coupled sets of polynomial
equations. In fact we would not expect {\em a-priori} that a
closed-form analytic solution would exist at all.

In the following section we will briefly describe the results of
\cite{DJ2}. Other studies of the close link between Seiberg-Witen
curves and random matrices include
\cite{FERRARI1,DGKV,GOPAKUMAR,SCHNITZER1,SCHNITZER2}. 

\section{Factorization solution of Seiberg-Witten curves for theories with
  fundamental matter} 

The main tool which allows to obtain the factorization of SW curves is
the study of the $\nn=2$ theory deformed by a tree-level superpotential
\eq
\label{e.tree}
W_{tree} = \sum_{p=1}^{N_c}  g_p \cdot \frac{1}{p} \tr \Phi^p.
\eqx
Then once the factorization solution $\uf_p$ is known, the {\em effective}
superpotential is given by
\eq 
\label{qep}
W_{eff} = \sum_{p=1}^{N_c} g_p \uf_p(\Lm,m_i,T),
\eqx
which should then be minimized with respect to the parameter $T$ of
$\uf_p$. If we integrate in $S$  by performing a Legendre
transformation with respect to $\log \Lambda^{2N_c-N_f}$ we obtain the
superpotential 
\eqn
\label{e.intin}
\!\!\!\!\!\!W_{eff} (S, u_1, \OM, \Lm) \!\!\!&=&\!\!\! S \log
\Lambda^{2N_c-N_f} +W_{eff}(S,u_1,\OM) = \nonumber\\
\!\!\!\!\!\!\!\!\!&=&\!\!\! S \log \Lambda^{2N_c-N_f} - S \log
\OM^{2N_c-N_f} \! +\! \sum_{p=1}^{N_c} g_p \uf_p (\OM, u_1) 
\eqnx
In order to obtain the factorization solution $\uf_p$ from random
matrix models we 
should first use the Dijkgraaf-Vafa prescription to obtain $W_{eff}(S)$ from
a random matrix expression and then recast it in the gauge-theoretic form 
$W_{eff}(S,u_1,\OM)$ defined by (\ref{e.intin}) which is {\em linear} in the
couplings $g_p$ --- this is the most
difficult part of the computation. Then one can directly
read off the factorization solution from the coefficients of the
couplings.    

The structure of the solution obtained in \cite{DJ2} is the following:
\eq
\label{e.ufpfinal}
\uf_p = N_c\, \upnc_p(R,T) +\sum_{i=1}^{N_f}\, \upm_p(R,T,m_i)
\eqx
where the two random matrix parameters $R$ and $T$ are 
related to the physical parameters $\Lm$, $u_1$ by the constraints
\eqn
\Lambda^{2N_c - N_f} &=& \frac{R^{N_c}}{\prod_{i=1}^{N_f}
       \f{1}{2}\left(m_i+T+\sqrt{(m_i+T)^2-4R}\right)} \\
u_1 &=& N_c T -\f{1}{2} \sum_{i=1}^{N_f} m_i+T-\sqrt{(m_i+T)^2 -4R}
\eqnx
In (\ref{e.ufpfinal}) $\upnc_p(R,T)$ is the factorization solution for
{\em pure} $\nn=2$ theory 
\eq
\upnc_p(R,T) = \f{1}{p}  \sum_{q=0}^{[p/2]} \bin{p}{2q} \bin{2q}{q}
R^{q} T^{p-2q} 
\eqx
while the explicit form of $\upm_p(R,T,m_i)$ is given in section 7 of
\cite{DJ2}. Here we just cite the result for $p\leq 3$:
\eqn
\upm_1(R,T,m)\!\! &=&\!\! \f{1}{2} \biggl( -m-T+\sqrt{(m+T)^2-4R} \biggr) \\
\upm_2(R,T,m)\!\! &=&\!\! \f{1}{4} \biggl( m^2-2R-T^2+(T-m) \sqrt{(m+T)^2-4R}
\biggr) \\
\upm_3(R,T,m)\!\! &=&\!\! \f{1}{6} \biggl(-m^3-6RT-T^3+ (m^2 +2R-mT+T^2) \cdot
  \nonumber \\
  && \;\;\; \cdot \sqrt{(m+T)^2-4R} \biggr)
\eqnx

Let us note some striking features of the solution
(\ref{e.ufpfinal}). Firstly, it has an extremely simple dependence on the
number of colours. Secondly, each flavour contributes
linearly. Thirdly, the solution for pure SYM theory appears as a part
of the expression. All of these features are quite surprising if we
keep in mind the very much nonlinear character of the factorization
problem for the curve (\ref{e.swgen}). 
Also from the physical point of
view such unexpected linearization seems to suggest some hidden
structure of the SYM theory with fundamental flavours.  

Recently, the factorization solution (\ref{e.ufpfinal}) was used in
\cite{YD} to rederive from the Seiberg-Witten curve perspective the
Affleck-Dine-Seiberg superpotential (\ref{e.ads}).

\section{Discussion}

The link of random matrices to supersymmetric gauge theories is one of
the most exciting theoretical developments in the last year. It is
notable as giving a completely new {\em exact} application of
random matrix models. The interest lies both in reinterpreting
classical random matrix results/ensembles in a new langage and
context, and in using the novel random matrix methods to obtain new
results in supersymmetric gauge theories. It raises also numerous
questions of a more mathematical nature. In particular it would be
interesting to understand the precise interrelation between
Calabi-Yau constructions and random matrices in the most general
setting. This link was in fact at the origin of the Dijkgraaf-Vafa
proposal. 

Last but not least it would be interesting to understand
the mathematical structures which allowed to use random matrix model
calculations to factorize Seiberg-Witen curves. The solution obtained
in \cite{DJ2} is based on a reinterpretation of a random matrix
formula using input from gauge-theoretical reasonings involving
such non-perturbative concepts as `integrating-in' etc. . It would be
fascinating to link directly the underlying geometry of factorization
of Seiberg-Witten curves with random matrix considerations. Another
interesting open problem would be to generalize the solution to the
case of non-complete factorization i.e. with at least one monopole
staying massive. This is of direct relevance to the study of the
global structure of $\nn=1$ vacua \cite{FERRARI2,PHASES1,PHASES2}.   

\bigskip

\vspace{12pt}

\noindent{\bf Acknowledgments.} I would like to thank Jan Ambj{\o}rn
and Yves Demasure with whom the results described here were
obtained. This work was supported in part by KBN grants 2P03B09622
(2002-2004), 2P03B08225 (2003-2006) and by the EU network on
``Discrete Random Geometry'', grant HPRN-CT-1999-00161. 

\vfill

\pagebreak

\end{document}